\documentclass[lineno]{biometrika}
\usepackage{titlesec}
\usepackage{amsmath}
\usepackage{amssymb}
\usepackage{subcaption}

\usepackage{hyperref}
\usepackage{bm}
\usepackage{newtxtext}
\usepackage[subscriptcorrection]{newtxmath}

\graphicspath{{./art/}}

\usepackage[plain,noend]{algorithm2e}

\makeatletter
\renewcommand{\algocf@captiontext}[2]{#1\algocf@typo. \AlCapFnt{}#2} 
\def\@algocf@capt@plain{top}
\renewcommand{\algocf@makecaption}[2]{%
  \addtolength{\hsize}{\algomargin}%
  \sbox\@tempboxa{\algocf@captiontext{#1}{#2}}%
  \ifdim\wd\@tempboxa >\hsize
    \hskip .5\algomargin%
    \parbox[t]{\hsize}{\algocf@captiontext{#1}{#2}}
  \else%
    \global\@minipagefalse%
    \hbox to\hsize{\box\@tempboxa}
  \fi%
  \addtolength{\hsize}{-\algomargin}%
}
\makeatother

\makeatletter
\def\ps@plain{\ps@empty}
\def\ps@headings{\ps@empty}
\let\ps@biom\ps@empty
\let\ps@biometrika\ps@empty
\makeatother

\begin{document}
\nolinenumbers

\jname{}
\jyear{}
\copyrightinfo{}
\makeatletter
\def\ps@plain{\ps@empty}
\makeatother




\title{Optimal Watermark Generation under Type I and Type II Errors}

\author{Hengzhi He}
\affil{Department of Statistics and Data Science, University of California, Los Angeles, 8125 Math Sciences Bldg, Los Angeles, CA, USA
\email{hengzhihe@g.ucla.edu}}
\author{Shirong Xu}
\affil{The Wang Yanan Institute for Studies in Economics, Xiamen University, Economics Building, Xiamen University Xiamen, 361005 China
\email{shirongxu5566@xmu.edu.cn}}
\author{Alexander Nemecek}
\affil{Department of Computer and Data Sciences, Case Western Reserve University,\\
Cleveland, OH 44106, USA
\email{ajn98@case.edu}}
\author{Jiping Li}
\affil{Department of Mathematics, University of California, Los Angeles, 520 Portola Plaza, MS 6363 Box 951555 Los Angeles, CA 90095-1555
\email{jipingli0324@g.ucla.edu}}
\author{Erman Ayday}
\affil{Department of Computer and Data Sciences, Case Western Reserve University,\\
Cleveland, OH 44106, USA
\email{exa208@case.edu}}
\author{Guang Cheng}
\affil{Department of Statistics and Data Science, University of California, Los Angeles, 8125 Math Sciences Bldg, Los Angeles, CA, USA
\email{guangcheng@ucla.edu}}

\markboth{H. He et~al.}{Biometrika style}

\maketitle

\begin{abstract}
Watermarking has recently emerged as a crucial tool for protecting the intellectual property of generative models and for distinguishing AI-generated content from human-generated data. Despite its practical success, most existing watermarking schemes are empirically driven and lack a theoretical understanding of the fundamental trade-off between detection power and generation fidelity. To address this gap, we formulate watermarking as a statistical hypothesis testing problem between a null distribution and its watermarked counterpart. Under explicit constraints on false-positive and false-negative rates, we derive a tight lower bound on the achievable fidelity loss, measured by a general $f$-divergence, and characterize the optimal watermarked distribution that attains this bound. We further develop a corresponding sampling rule that provides an optimal mechanism for inserting watermarks with minimal fidelity distortion. Our result establishes a simple yet broadly applicable principle linking hypothesis testing, information divergence, and watermark generation.
\end{abstract}


\section{Introduction}
Watermarking has recently become increasingly popular across various domains, being applied to tabular data (\cite{fang2025muse,he2024watermarking,karthikeyan2025hashmark}), image (\cite{wen2024tree,fernandez2023stable,gunnundetectable}) and text (\cite{kirchenbauer2023watermark,zhao2023provable,aaronson2024slides,kuditipudirobust,nemecek2024topic,li2025robust,li2025statistical,xie2025debiasing,dathathri2024scalable,giboulot2024watermax}) to distinguish AI-generated content from human creations and to protect model ownership.

However, watermarking faces substantial challenges in real-world deployment. Beyond concerns about robustness (\cite{zhang2023watermarks}) and forgery risk (\cite{gulearnability,sadasivan2025can}), perhaps the most critical factor limiting industry adoption is the possibility that watermarking degrades model quality (\cite{liu2025position}). For many users, any noticeable degradation in output quality will lead them to prefer unwatermarked alternatives; if watermarking causes the generated output to deviate from the model's natural behavior, users may simply switch to competing generative models. As a result, quality degradation becomes a central barrier to practical deployment.

This leads to a natural question: if a watermarking scheme is required to satisfy prescribed false-positive and false-negative rates, what is the {minimum} fidelity loss that must be incurred, and how can a watermarked distribution attaining this optimal trade-off be constructed? The study of this problem remains relatively preliminary from a statistical perspective, with only a few recent 
papers (\cite{cai2024towards,he2025theoretically}) examining aspects of this question (see the Appendix~\ref{additiaonal_related_works} for a detailed discussion of related work).

In this work, we abstract away the specifics of any particular watermarking implementation and formulate the problem purely as a binary hypothesis test between a null distribution and its watermarked counterpart~\citep{li2025statistical}. Under fixed Type~I and Type~II error constraints, we derive a tight lower bound on the achievable fidelity loss, measured by a general $f$-divergence, and characterize the optimal distribution that achieves this infimum. We further show that this optimal distribution can be realized by reinforcement learning or simple rejection-sampling.

\section{Problem formulation}

We formalize the watermarking task as a constrained distributional modification problem. Let $F$ denote the original, unwatermarked distribution, and let $G$ be a candidate watermarked distribution. A detector is represented by a binary decision rule $D:\mathcal{X}\to{0,1}$, where $D(x)=1$ indicates that the datum $x$ is declared to contain a watermark. For any distribution $P$ (e.g., $F$ or $G$) and any measurable set $S$, we use $P(S)$ to denote the probability that a sample drawn from $P$ falls in $S$.

To ensure detectability, we impose constraints on the false-positive and false-negative rates:
\begin{align}
\label{eq:FPR}
F(\{D(x)=1\}) &\le \alpha,\\[3pt]
\label{eq:FNR}
G(\{D(x)=1\}) &\ge 1-\beta,
\end{align}
for prescribed levels $\alpha,\beta\in[0,1]$.  
Here,~\eqref{eq:FPR} limits the probability of falsely flagging unwatermarked content (e.g., human-written text) as watermarked, while~\eqref{eq:FNR} ensures that watermarked samples are detected with high probability and are not frequently missed.

Focusing on fidelity, the objective can be interpreted as seeking a distribution $G$ that satisfies the detection constraints~\eqref{eq:FPR} and~\eqref{eq:FNR}, so that watermarked content remains distinguishable from unwatermarked samples, while deviating from the original distribution $F$ as little as possible. To quantify the discrepancy between $F$ and $G$, we employ the $f$-divergence, a general framework that {covers a broad family of metrics} used to measure distributional differences in modern image (\cite{nowozin2016f}) and text (\cite{go2023aligning}) generation tasks. The $f$-divergence is defined as:
$
D_f(G\|F) = \int f\!\left(\frac{dG}{dF}\right)\, dF,
$
where $f:\mathbb{R}_{+}\to\mathbb{R}$ is a convex function with $f(1)=0$.

The $f$-divergence is particularly significant because it unifies many discrepancy measures that are routinely used across different data modalities. 
For example, the {total variation distance} is a special case of the $f$-divergence (see example~\ref{example_total}) and is widely used as a fidelity criterion for tabular and other structured data (\cite{zhu2025tabwak,lautrup2025syntheval}). In the image domain, $f$-divergences also play a central role. 
Many popular generative adversarial networks are trained by minimizing an $f$-divergence~\citep{nowozin2016f}; 
Moreover, one of the most important special cases of 
$f$-divergences, the Kullback--Leibler (KL) divergence (see example~\ref{example_kl}), lies at the heart of modern diffusion models, since training diffusion models can be seen as minimizing a variational upper bound of KL divergence (\cite{ho2020denoising}). In the context of large language models, the KL divergence is also particularly relevant: the standard {Perplexity} metric for evaluating text generation quality is essentially a monotone transformation of KL when the data distribution is fixed~\citep{manning1999foundations}. 

Under our formulation, designing an optimal watermarked distribution $G$ reduces to solving
\begin{align}
\label{target}
G^\star 
= \arg\min_{G:\,\exists\,D\;\text{satisfying}\;\eqref{eq:FPR}\text{--}\eqref{eq:FNR}}
D_f(G\|F),
\end{align}
which makes explicit the best achievable fidelity-detection trade-off under the prescribed error constraints. This formulation abstracts away the implementation details of any specific watermarking scheme and isolates the fundamental trade-off dictated solely by hypothesis-testing constraints.  
As we show in the next section, the problem admits a clean and tractable solution: the optimal $G$ has an explicit closed form, and it can be realized via a reinforcement learning based optimization or a simple rejection-sampling mechanism.

\section{Optimal Watermarked Distribution and Fidelity Loss lower bound}
In this section, we solve the optimization problem in~\eqref{target} and characterize both the minimal achievable fidelity loss and the optimal watermarked distribution that attains it.

\begin{theorem}[Optimal watermarked distribution and minimal $f$-divergence]
\label{thm:f-optimal}
Suppose $F$ is the original distribution, and let a watermarked distribution $G$ together with a detector $D:\mathcal{X}\to \{0,1\}$ satisfy the error constraints~\eqref{eq:FPR} and~\eqref{eq:FNR} for some $\alpha,\beta\in(0,1)$ with $\alpha \le 1-\beta$. Let $S=\{x\in\mathcal{X}: D(x)=1\}$ denote the detection region. For any convex function $f:\mathbb{R}_{+}\to\mathbb{R}$ with $f(1)=0$, $G$ satisfies the following lower bound:
\begin{equation}
\label{eq:f-lb-corner}
D_f(G\|F)
\;\ge\;
\alpha\, f\!\Big(\frac{1-\beta}{\alpha}\Big)
+
(1-\alpha)\, f\!\Big(\frac{\beta}{1-\alpha}\Big).
\end{equation}
Moreover, the bound in~\eqref{eq:f-lb-corner} is tight and achieved by 
\begin{equation}
\label{eq:G-star-final}
G^\star(A)
=
\frac{1-\beta}{\alpha}\,F(A\cap S)
+
\frac{\beta}{1-\alpha}\,F(A\cap S^c),
\qquad A\in\mathrm{Borel}(\mathcal{X}),
\end{equation}
where $S=\{x:D(x)=1\}$ is chosen so that $F(S)=\alpha$.  
Then $G^\star$ satisfies $G^{\star}(\{D(x)=1\})=1-\beta$ and
\begin{equation}
\label{eq:f-lb-value}
D_f(G^\star\|F)
=
\alpha\, f\!\Big(\frac{1-\beta}{\alpha}\Big)
+
(1-\alpha)\, f\!\Big(\frac{\beta}{1-\alpha}\Big).
\end{equation}
\end{theorem}

To illustrate the implications of Theorem~\ref{thm:f-optimal} for concrete choices of $f$, we next provide an example corresponding to the KL divergence. We also include two examples corresponding to the total variation divergence and Chi-square divergence in the Appendix \ref{additiaonal_related_works}.

\begin{example}[Kullback--Leibler divergence]
\label{example_kl}
When the generating function is chosen as
$
f_{\mathrm{KL}}(t)= t\log t,
$
the resulting $f$-divergence coincides with the KL divergence,
$
D_{\mathrm{KL}}(G\|F)
=\int \frac{dG}{dF}\log\!\left(\frac{dG}{dF}\right)\, dF.
$ Substituting $f_{\mathrm{KL}}$ into the lower bound~\eqref{eq:f-lb-corner} yields
$
D_{\mathrm{KL}}(G\|F)
\;\ge\;
\alpha\Big(\tfrac{1-\beta}{\alpha}\log\!\Big(\tfrac{1-\beta}{\alpha}\Big)\Big)
+
(1-\alpha)\Big(\tfrac{\beta}{1-\alpha}\log\!\Big(\tfrac{\beta}{1-\alpha}\Big)\Big).
$
This simplifies to the more compact form
$
D_{\mathrm{KL}}(G\|F)
\;\ge\;
(1-\beta)\log\!\Big(\frac{1-\beta}{\alpha}\Big)
+
\beta\log\!\Big(\frac{\beta}{1-\alpha}\Big).
$
\end{example}

\begin{remark}
A recent work derives an $f$-divergence bound with the same analytical
form as Theorem~\ref{thm:f-optimal}, but from a completely different
motivation. Their result does not establish tightness of this bound for
general $f$, nor does it address the optimization problem in~\eqref{target} or provide a constructive procedure for obtaining a
distribution that attains the bound. In contrast, our work derives the
optimal value of~\eqref{target} together with its achieving
distribution. See Appendix~\ref{additiaonal_related_works} for further
discussion.
\end{remark}

\section{Constructing the watermarked generator}

The optimal distribution $G^\star$ in Theorem~\ref{thm:f-optimal} describes the best possible watermarking approach  to preserve fidelity under fixed Type~I and Type~II error constraints.  
In practice, however, the way we realize this ideal distribution depends on how much control we have over the underlying generator.

If we are able to fine-tune or retrain the model, then we can learn a watermarked generator whose output distribution is driven toward $G^\star$ by optimizing a suitable objective.  
This leads naturally to a reinforcement learning (RL)-style formulation, in which we can parameterize a policy and train it using a proximal policy optimization algorithm (\cite{schulman2017proximal}) so that the induced sampling distribution matches the theoretical optimum.

On the other hand, if the base generator cannot be modified, which is often the case for large closed-source models, we can still implement watermarking by adjusting the sampling procedure alone.  
Leveraging the closed-form density ratio $dG^\star/dF$ given by Theorem~\ref{thm:f-optimal}, we can construct a simple acceptance-rejection sampler that transforms base samples from $F$ into exact samples from $G^\star$ without changing the model's parameters.

We describe these two approaches in turn, beginning with the learning-based formulation.

\begin{theorem}[RL objective recovers the optimal watermarked distribution]
\label{thm:rl-optimal}
Let $F$ be the original distribution and $D:\mathcal{X}\to\{0,1\}$ a detector with detection region $S=\{x:D(x)=1\}$ and $F(S)=\alpha$.
Fix a target Type~II error level $\beta\in[0,1]$ with $\alpha\le 1-\beta$, and let $G^\star$ be the optimal watermarked distribution from Theorem~\ref{thm:f-optimal}, i.e.
\[
\frac{dG^\star}{dF}(x)
=
\begin{cases}
\dfrac{1-\beta}{\alpha}, & x\in S,\\[6pt]
\dfrac{\beta}{1-\alpha}, & x\in S^c.
\end{cases}
\]
Consider the KL-regularized RL objective
\begin{equation}
\label{eq:rl-objective}
J(\pi)
=
\mathbb{E}_{X\sim\pi}\Big[r(X) - \log\Big(\frac{\pi(X)}{F(X)}\Big)\Big],
\end{equation}
defined over all probability measures $\pi$ that are absolutely continuous with respect to $F$.
Define the reward
$
r(x) = A\,D(x),
\
A := \log\!\Big(\frac{(1-\beta)(1-\alpha)}{\alpha\beta}\Big).
$
Then $J(\pi)$ has a unique maximizer, and this maximizer coincides with $G^\star$:
$
\arg\max_\pi J(\pi) = G^\star.
$
In particular, any RL procedure that maximizes \eqref{eq:rl-objective} over a rich enough policy class $\{\pi_\phi\}$ containing $G^\star$ and converges to a global maximizer will learn the optimal watermarked distribution $G^\star$.
\end{theorem}
The result in Theorem \ref{thm:rl-optimal} allows us to directly optimize $J(\pi_\phi)$ using modern policy-gradient methods. 
In practice, we can parameterize $\pi_\phi$ by a model initialized at the base model and employ standard proximal policy optimization to maximize $J(\pi_\phi)$.  
The resulting fine-tuned policy produces samples whose distribution approaches the theoretical optimum $G^\star$.
\begin{remark}
A recent empirical line of work studies model-level watermarking for language models via RL (see \cite{xu2024learning}). In their approach, the detector $D$ is implemented as a discriminative
classifier on prompt--response pairs, and the generator is fine-tuned to {increase} the detector score while staying close to the
base model through a KL regularization term.
Consequently, the reward used for training the generator is essentially the detector output minus a KL penalty, and the relative weighting of these components is selected empirically. Such a design does not provide any guarantee on the achievable false-positive and false-negative rates or on the resulting fidelity loss.

In contrast, our framework starts from a formal hypothesis-testing
formulation and explicitly derives the optimal watermarked distribution
$G^\star$ under prescribed Type~I and Type~II error constraints.
Moreover, Theorem~\ref{thm:rl-optimal} shows how to construct a {principled} reward function whose unique maximizer is precisely
$G^\star$, rather than tuning heuristic reward weights.
Thus our results characterize both the optimal fidelity-detection
trade-off and an RL objective that provably recovers the optimal
watermarked distribution.
\end{remark}

While Theorem~\ref{thm:rl-optimal} shows that the optimal distribution
$G^\star$ can be learned by a suitable RL objective
when the model can be fine-tuned, one might argue that this assumption may be unrealistic in
many applications. Some models often provide only a sampling interface, making
parameter-level training infeasible.
However, in such settings, the closed-form likelihood ratio $dG^\star/dF$ from
Theorem~\ref{thm:f-optimal} offers a direct alternative: it allows samples
from $F$ to be converted into exact samples from $G^\star$ without
modifying the model.
This leads to a simple, tuning-free rejection-sampling scheme, given in
Algorithm~\ref{alg:two_rate_sampling_pvalue} and justified by
Theorem~\ref{thm:sampling-optimal}.

\begin{algorithm}
\caption{Two-Rate Acceptance Sampling}
\label{alg:two_rate_sampling_pvalue}

\KwIn{sampler $x\sim F$, detector $D(x)\in\{0,1\}$, FPR $\alpha$, FNR $\beta$}

Compute the two importance weights
\[
w_1=\frac{1-\beta}{\alpha},
\qquad
w_0=\frac{\beta}{\,1-\alpha\,}.
\]

\Repeat{a sample is accepted}{
    Draw $x\sim F$.\;
    \uIf{$D(x)=1$}{
        Accept $x$ with probability $1$.\;
    }
    \Else{
        Accept $x$ with probability $w_0/w_1$.\;
    }
}

\KwOut{accepted sample $x^\star$}

\end{algorithm}

\begin{theorem}[Sampling-based construction of the optimal watermarked distribution]
\label{thm:sampling-optimal}
Let $F$ be the original distribution and $D:\mathcal{X}\to\{0,1\}$ a detector with detection region
$
S=\{x: D(x)=1\}, \qquad F(S)=\alpha.
$

Then the distribution of the accepted sample by Algorithm \ref{alg:two_rate_sampling_pvalue} is exactly the optimal watermarked distribution $G^\star$ from Theorem~\ref{thm:f-optimal}, i.e.
$
G^\star(A)
=
w_1\,F(A\cap S)
+
w_0\,F(A\cap S^c),
\qquad
\forall A.
$
\end{theorem}

\begin{remark}
Algorithm~\ref{alg:two_rate_sampling_pvalue} is closely related in spirit to the multi-sample
selection strategy used in~\cite{giboulot2024watermax}, but there is an essential
difference in the sampling mechanism.

Our construction performs {true rejection sampling} with proposal distribution
$F$ and target distribution $G^\star$: at each step we draw $x\sim F$ and accept it
with a probability proportional to the optimal density ratio $dG^\star/dF(x)$, so
that the law of the accepted sample coincides {exactly} with $G^\star$ as
characterized in Theorem~\ref{thm:f-optimal}.
In the special case $\beta=0$, this reduces to pure conditioning on the detection
region, i.e. $G^\star(\cdot)=F(\cdot\mid S)$.

In contrast,~\cite{giboulot2024watermax} adopts a {finite multiple-sampling} scheme: for each query, it generates a fixed finite batch of $n$ independent candidates from $F$ and
selects the one with the smallest $p$-value. The resulting output distribution is
stochastically biased toward lower $p$-values and can be viewed as an approximation
to $G^\star$, but for any finite $n$ it does not coincide with the exact target
distribution.
\end{remark}

\section{Experiments}

We demonstrate the effectiveness of both the RL-based and rejection-sampling watermarking generators on a real tabular dataset. We use the National Poll on Healthy Aging dataset (\cite{malani2019npha}), treat each row as a distinct state $x$, and take the empirical distribution that assigns mass $1/N$ to each row as the ground-truth distribution $F$.
We construct an unsupervised detector by randomly assigning half of the rows to label 0 and half to label 1, then fitting a gradient-boosting classifier on all covariates.
The fitted classifier yields a score function $s(x)$, and for several thresholds $\tau$ we define $D_\tau(x)=\mathbf{1}\{s(x)\ge\tau\}$ and calculate the false-positive rate
$\alpha(\tau)=\mathbb{P}_{X \sim F}\{D_\tau(X)=1\}$ in the ground-truth distribution.

For each $(\tau,\beta)$ satisfying the feasibility condition $\alpha(\tau) < 1-\beta$, we construct a watermarked distribution in two ways.
First, we form the \emph{rejection-sampling generator} $\widehat{G}_{S}$ of Algorithm~\ref{alg:two_rate_sampling_pvalue};
Secondly, we train an \emph{RL-based generator} by maximizing the KL-regularized objective
$
J(\pi)
=
\mathbb{E}_{X\sim\pi}[r_\tau(X)]
-
KL(\pi\|F),
$
where $\pi$ is a tabular softmax policy over all possible states and $r_\tau$ is chosen as in Theorem~\ref{thm:rl-optimal} so that the unique maximizer of $J(\pi)$ coincides with the optimal distribution $G^\star$ from Theorem~\ref{thm:rl-optimal} associated with $(\alpha(\tau),\beta)$.
Because the state space is finite, the resulting policy $\hat\pi$ induces a discrete distribution $\widehat G_{R}$. For both $\widehat{G}_{S}$ and $\widehat{G}_{R}$, we compute the KL divergence $KL(\hat G\|F)$ and compare it with the theoretical lower bound
$
L(\alpha,\beta)
=
(1-\beta)\log\frac{1-\beta}{\alpha}
+
\beta\log\frac{\beta}{1-\alpha}.
$
\begin{figure}[t]
    \centering
    \begin{subfigure}{0.48\textwidth}
        \centering
        \includegraphics[width=\textwidth]{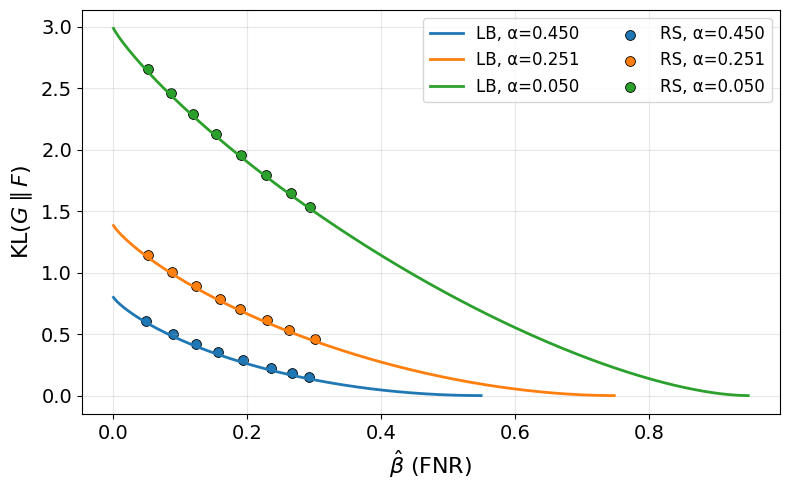}
        \caption{Rejection-sampling generator.}
        \label{fig:rs}
    \end{subfigure}
    \hfill
    \begin{subfigure}{0.48\textwidth}
        \centering
        \includegraphics[width=\textwidth]{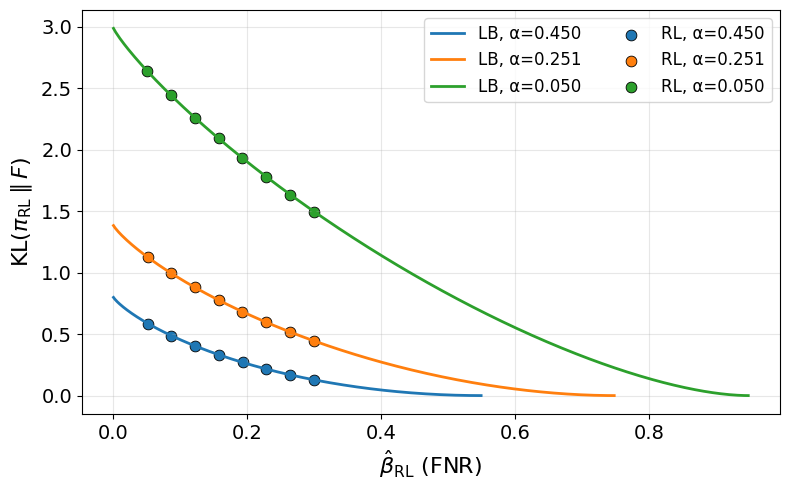}
        \caption{RL-based generator.}
        \label{fig:rl}
    \end{subfigure}

    \vspace{0.8em}

    \caption{Empirical KL divergence versus theoretical lower bound for the three watermarking generators.}
    \label{fig:comparison}
\end{figure}

As shown in Figure~\ref{fig:comparison}, both methods yield watermarked distributions whose KL divergence from the original distribution approaches the theoretical lower bound. We also implement the sampling algorithm proposed in \cite{giboulot2024watermax} and find that it does not achieve the optimal fidelity-detectability trade-off; see Appendix~\ref{additional_experiment} for details.

\section{Conclusion}In this work, we investigate the problem of optimal watermark insertion \textit{under any given detector}. Once the detector is fixed and the Type~I and Type~II error rates are prescribed, we fully characterize the best achievable fidelity loss. In particular, we derive a tight lower bound on the required fidelity loss for a general $f$-divergence, and explicitly construct the optimal watermarked distribution that attains this bound. Our results show that, regardless of how the detector is designed, there exists an optimal way to watermark the base distribution so as to satisfy the detection constraints while incurring the minimal possible distortion. Building on this theoretical insight, we further provide two algorithmic realizations of the optimal watermarked distribution: (i) an RL-based optimization procedure for settings where the generator can be fine-tuned, and (ii) a lightweight acceptance-rejection sampler that applies even when only black-box sampling access to the base model is available. Together, these results establish a principled and broadly applicable framework for watermark generation under arbitrary detectors.

\section*{Declaration of the use of generative AI and AI-assisted technologies}
During the preparation of this work the author(s) used GPT 5.1 and Google Gemini to polish the writing and improve the clarity of content delivery. After using this tool/service the author(s) reviewed and edited the content as necessary and take(s) full responsibility for the content of the publication.


\label{SM}

\bibliographystyle{biometrika}
\bibliography{Bibliography-MM-MC}


\clearpage

\appendix

\appendixone
\section{Additional Discussions on Related Works}
\label{additiaonal_related_works}
Watermarking has recently attracted substantial attention in the statistics and machine learning communities. 
Beyond the development of new watermarking algorithms, several works have begun to investigate the fundamental
trade-off between fidelity and detectability, aiming to understand the statistical limits of how much a distribution
can be perturbed while still remaining reliably detectable. (\cite{cai2024towards,he2025theoretically})

Among these works, the study most closely related to ours is the recent paper by~\cite{cai2024towards}. 
They formalize the watermarking problem as an optimization task that seeks to {minimize the KL divergence} 
between a watermark model distribution and the original model distribution, subject to a constraint on the 
{increase in the probability of the green-list}, which serves as a proxy of the detection power. Compared with their formulation, our framework is substantially more general. 
Rather than focusing on the specific green-red list watermarking mechanism, 
we model watermarking as a distributional modification problem applicable to 
{any} watermarking scheme and {any} data modality. 
Within this broader framework, we derive the exact optimal 
fidelity-detectability trade-off under explicit false-positive and 
false-negative error constraints, which is strictly tighter than the information-theoretic 
distortion bound established in Proposition~3.1 of~\cite{cai2024towards}. 

To be more specific, the proposition~3.1 in~\cite{cai2024towards} is equivalent to 

\begin{equation}
\label{eq:cai-lower-bound}
    D_{\mathrm{KL}}(G\,\|\,F)
    \;\ge\;  
    -\log\!\bigl((\alpha+\beta)(2-\alpha-\beta)\bigr) \stackrel{\triangle}{=}
 g_1(\alpha,\beta).
\end{equation}

In contrast, our framework yields the following \emph{exact} and \emph{attainable} optimal distortion bound:
\begin{equation}
\label{our_bound}
D_{\mathrm{KL}}(G\,\|\,F)\geq\;
(1-\beta)\log\!\Bigl(\frac{1-\beta}{\alpha}\Bigr)
\;+\;
\beta\log\!\Bigl(\frac{\beta}{1-\alpha}\Bigr) \stackrel{\triangle}{=}
 g_2(\alpha,\beta)
\end{equation}
which is achieved by an explicit optimal construction of the watermarked distribution $G^\star$. It is easy to check that $g_2(\alpha,\beta)>g_1(\alpha,\beta)$ for any $0<\alpha,\beta<1$ with $\alpha+\beta<1$. Therefore our bound is strictly tighter than the bound obtained in \cite{cai2024towards}.

\cite{he2025theoretically} is another relevant line of work.
Their goal is to jointly optimize both the watermarking distribution and the detector,
under constraints on the false-positive probability and on a distortion budget measured
between the original large language model (LLM) distribution and the watermarked one.
A key feature of their formulation is the introduction of an \emph{auxiliary random sequence}
$\zeta_{1:T}$, which is available to both the watermarking encoder and the detector.
The optimal detector in their framework critically relies on exploiting the statistical
dependence between $(X_{1:T},\zeta_{1:T})$ under watermarking and the independence structure
under the null model, where $X_{1:T}$ denotes the text.

While mathematically elegant, this modeling assumption has important practical limitations.
In real deployments, such an auxiliary sequence $\zeta_{1:T}$ would necessarily be generated by a pseudorandom function seeded by the key, and is therefore {deterministic}.
Thus the transition from true randomness to pseudorandomness inevitably incurs additional fidelity loss that their theory does not characterize.
Moreover, the entire construction fundamentally relies on the token-by-token sampling structure of LLMs and does not extend to other data modalities such as tabular or image generation. 

Similar auxiliary-sequence assumptions also appear in several recent LLM watermarking papers (\cite{li2025statistical,xie2025debiasing}).
This key structural difference further distinguishes our work from these approaches: in contrast to theoretical results that rely on idealized auxiliary randomness, our fidelity analysis is strict and {already incorporates the potential pseudorandomness-induced distortion that inevitably arises in practical watermarking systems, and therefore captures the true operational cost of watermark embedding.

Besides, a recent mathematical paper~\cite{mullhaupt2025bounding} derives an $f$-divergence bound with
the same analytic form as ours in Theorem~\ref{thm:f-optimal}, but the
problem studied there is quite different. Their focus is on the geometry
of the Neyman--Pearson region, namely the set of all achievable
$(\alpha,\beta)$ pairs for testing. However, they do not establish
tightness of the bound for general $f$; their only tightness result is
for the hockey-stick divergence (see Example~1 in their paper).  

In comparison, our motivation is different and so are the conclusions.
We prove that the bound is tight for \emph{every} $f$-divergence, and,
for any baseline distribution $F$ and any feasible pair
$(\alpha,\beta)$, we explicitly construct the achieving distribution
$G^\star$ that attains the optimal value. This constructive achievability
is not addressed in~\cite{mullhaupt2025bounding}. Moreover, to the best of our knowledge, our paper is the first to apply
such tight $f$-divergence optimality results to generative watermarking.

\section{Additional Experiment Results}
\label{additional_experiment}
In addition to the two theoretically optimal generators, we also implement the 
``best-of-$m$'' sampling strategy used in prior work (\cite{giboulot2024watermax}) on the same  National Poll on Healthy Aging dataset, 
which draws $m$ i.i.d. candidates from $F$ and selects the one with the highest 
detector score. 
As shown in Figure~\ref{fig:best_of_m}, this heuristic does not reach the optimal 
fidelity-detectability trade-off and remains above our theoretical 
lower bound.

\begin{figure}
    \centering
    \includegraphics[width=0.7\linewidth]{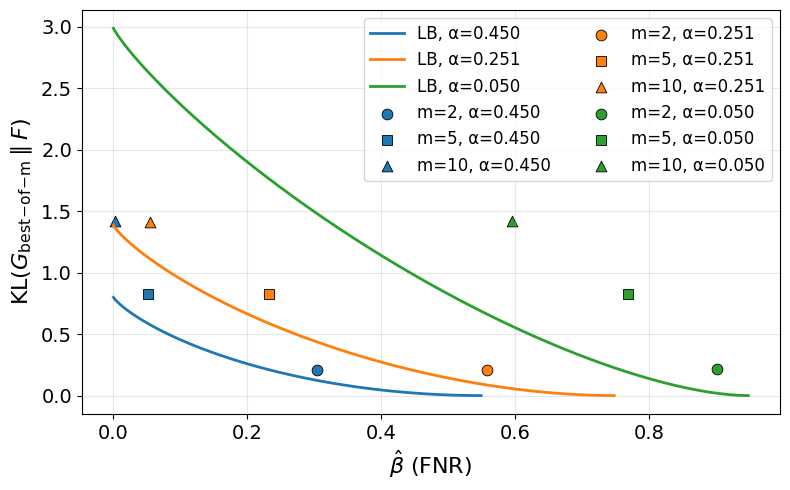}
    \caption{Empirical KL divergence of the best-of-$m$ sampling generator 
(implemented by drawing $m$ candidate samples from $F$ and selecting the 
one with the highest detector score), compared against the optimal 
fidelity-detectability lower bound.}

    \label{fig:best_of_m}
\end{figure}

\section{Additional Examples for Theorem \ref{thm:f-optimal}}
\label{additional_example}
In this section, we provide two more examples for Theorem~\ref{thm:f-optimal}.
\begin{example}[Total variation divergence]
\label{example_total}
When the generating function is chosen as
$
f_{\mathrm{TV}}(t)=\tfrac12 |t-1|,
$
the resulting $f$-divergence coincides with the total variation distance
$
D_{\mathrm{TV}}(G\|F)
=\frac12\int \bigg|\frac{dG}{dF}-1\bigg|\, dF=\frac12\int \bigg|{dG}-{dF}\bigg|\,.
$ Substituting $f_{\mathrm{TV}}$ into the lower bound~\eqref{eq:f-lb-corner} yields
$
D_{\mathrm{TV}}(G\|F)
\;\ge\;
\frac12\,\alpha\Big|\frac{1-\beta}{\alpha}-1\Big|
+
\frac12\,(1-\alpha)\Big|\frac{\beta}{1-\alpha}-1\Big|.
$
Under the feasibility condition $\alpha\le 1-\beta$, we have
$\frac{1-\beta}{\alpha}\ge 1$ and $\frac{\beta}{1-\alpha}\le 1$, so the bound simplifies to
$
D_{\mathrm{TV}}(G\|F)\;\ge\;1-\alpha-\beta.
$
Thus Theorem~\ref{thm:f-optimal} recovers the classical total-variation lower bound in
binary hypothesis testing.
\end{example}

\begin{example}[Chi-square divergence]
When the generating function is chosen as
$
f_{\chi^2}(t) = (t-1)^2,
$
the resulting $f$-divergence coincides with the Pearson chi-square divergence,
$
D_{\chi^2}(G\|F)
=
\int \Big(\frac{dG}{dF}-1\Big)^2\, dF.
$ Substituting $f_{\chi^2}$ into the lower bound~\eqref{eq:f-lb-corner} yields
$
D_{\chi^2}(G\|F)
\;\ge\;
\alpha\Big(\frac{1-\beta}{\alpha}-1\Big)^2
+
(1-\alpha)\Big(\frac{\beta}{1-\alpha}-1\Big)^2=\frac{(1-\alpha-\beta)^2}{\alpha(1-\alpha)}.
$
\end{example}

\section{Proof of Theorems}

In this section, we provide the proofs of the main theorems.

\par
\noindent\textbf{Proof of Theorem~\ref{thm:f-optimal}.} Let $S=\{x:D(x)=1\}$ be the detection region of a feasible detector $D$ and set
\[
a = F(S),\qquad q = G(S).
\]
The constraints \eqref{eq:FPR} and \eqref{eq:FNR} imply
\[
0 < a \le \alpha,
\qquad
1-\beta \le q \le 1
\]
.

\medskip
\noindent{Step 1: Coarse-graining onto $\{S,S^c\}$.}
Consider the measurable map $\pi:\mathcal{X}\to\{1,0\}$ given by $\pi(x)=\mathbf{1}\{x\in S\}$, and let
\[
F_\pi = F\circ\pi^{-1} = \mathrm{Bernoulli}(a),
\qquad
G_\pi = G\circ\pi^{-1} = \mathrm{Bernoulli}(q).
\]
By the data-processing inequality for $f$-divergences (\cite{polyanskiy2025information}),
\[
D_f(G\|F) \;\ge\; D_f(G_\pi\|F_\pi).
\]
For these Bernoulli distributions we have
\[
D_f(G_\pi\|F_\pi)
=
a\,f\!\Big(\frac{q}{a}\Big)
+
(1-a)\,f\!\Big(\frac{1-q}{1-a}\Big)
=:\phi(a,q).
\]
Thus
\[
D_f(G\|F) \;\ge\; \phi(a,q),
\]
for some $(a,q)$ in the following feasible rectangle
\[
\mathcal{R}_{\alpha,\beta}
=
\{(a,q): 0<a\le\alpha,\ 1-\beta\le q\le 1\}.
\]

\medskip
\noindent{Step 2: Minimization over $q$ for fixed $a$.}
Fix $a\in(0,\alpha]$ and view $\phi(a,q)$ as a function of $q$ on $[1-\beta,1]$.  
Writing $t_1=q/a$ and $t_2=(1-q)/(1-a)$, we obtain
\[
\frac{\partial\phi}{\partial q}(a,q)
=
f'(t_1) - f'(t_2).
\]
Since $f$ is convex, $f'$ is monotone nondecreasing on $(0,\infty)$, and
\[
t_1 \ge t_2
\iff
\frac{q}{a} \ge \frac{1-q}{1-a}
\iff
q \ge a.
\]
On the feasible set we have $q\ge 1-\beta \ge \alpha \ge a$, hence $t_1\ge t_2$ and
\[
\frac{\partial\phi}{\partial q}(a,q) \ge 0
\quad\text{for all }q\in[1-\beta,1].
\]
Thus, for each fixed $a$,
\[
\min_{1-\beta\le q\le 1} \phi(a,q)
= \phi(a,1-\beta).
\]

\medskip
\noindent{Step 3: Minimization over $a$.}
Define
\[
\psi(a)
:=
\phi(a,1-\beta)
=
D_f\big(\mathrm{Bernoulli}(1-\beta)\,\big\|\,\mathrm{Bernoulli}(a)\big),
\qquad a\in(0,\alpha].
\]
For fixed $P$, the map $Q\mapsto D_f(P\|Q)$ is convex in $Q$ (\cite{polyanskiy2025information}), and achieves its unique minimum at $Q=P$, where the divergence is zero.  
In the Bernoulli case this means that $\psi(a)$ is convex in $a$ and minimized at $a=1-\beta$.  
Our feasible region satisfies $0<a\le\alpha\le 1-\beta$, so on $(0,\alpha]$ we are to the left of the minimizer and convexity forces $\psi$ to be nonincreasing.  
Hence
\[
\min_{0<a\le\alpha} \psi(a)
= \psi(\alpha)
=
\alpha\, f\!\Big(\frac{1-\beta}{\alpha}\Big)
+
(1-\alpha)\, f\!\Big(\frac{\beta}{1-\alpha}\Big).
\]
Combining the three steps, we obtain the lower bound~\eqref{eq:f-lb-corner}.

\medskip
\noindent{Step 4: Attaining the bound.}
Let $S$ be chosen so that $F(S)=\alpha$, and define $G^\star$ by~\eqref{eq:G-star-final}.  
Then
\[
G^\star(S)
=
\frac{1-\beta}{\alpha}F(S)
=
1-\beta,
\qquad
G^\star(S^c)
=
\frac{\beta}{1-\alpha}F(S^c)
=
\beta,
\]
so~\eqref{eq:FPR} and~\eqref{eq:FNR} hold with equality, and
\[
\frac{dG^\star}{dF}(x)
=
\begin{cases}
\dfrac{1-\beta}{\alpha}, & x\in S,\\[6pt]
\dfrac{\beta}{1-\alpha}, & x\in S^c.
\end{cases}
\]
Substituting this likelihood ratio into the definition of $D_f(G^\star\|F)$ gives~\eqref{eq:f-lb-value}, showing that the lower bound is attained. 

\noindent\textbf{Proof of Theorem~\ref{thm:rl-optimal}.} By Donsker and Varadhan's variational formula (\cite{donsker1983asymptotic}), for any choice of $r$, the unique maximizer of~\eqref{eq:rl-objective} is given by 
\[
\frac{dH_r}{dF}(x) \propto e^{r(x)}.
\]

Now take $r(x) = A D(x)$ with $A$ as in the theorem statement.  
Then $r(x)$ is constant on $S$ and $S^c$, and
\[
e^{r(x)}
=
\begin{cases}
e^{A}, & x\in S,\\[3pt]
1, & x\in S^c.
\end{cases}
\]
Hence
\[
\frac{dH_r}{dF}(x)
=
\begin{cases}
c_1, & x\in S,\\[3pt]
c_0, & x\in S^c,
\end{cases}
\]
for some positive constants $c_1,c_0$ determined by normalization.
Indeed,
\[
H_r(S)
=
\frac{e^{A}F(S)}{e^{A}F(S)+F(S^c)}
=
\frac{e^{A}\alpha}{e^{A}\alpha + 1-\alpha}.
\]
With
\[
A = \log\!\Big(\frac{(1-\beta)(1-\alpha)}{\alpha\beta}\Big),
\]
we have $e^{A}\alpha\beta = (1-\beta)(1-\alpha)$ and a direct substitution shows that $H_r(S)=1-\beta$ and $H_r(S^c)=\beta$.
Writing $H_r(S)=\alpha c_1$ and $H_r(S^c)=(1-\alpha)c_0$, we obtain
\[
\alpha c_1 = 1-\beta,
\qquad
(1-\alpha)c_0 = \beta,
\]
so that
\[
c_1=\frac{1-\beta}{\alpha},
\qquad
c_0=\frac{\beta}{1-\alpha}.
\]
Thus
\[
\frac{dH_r}{dF}(x)
=
\begin{cases}
\dfrac{1-\beta}{\alpha}, & x\in S,\\[6pt]
\dfrac{\beta}{1-\alpha}, & x\in S^c,
\end{cases}
\]
which is exactly the density ratio $dG^\star/dF$ from Theorem~\ref{thm:f-optimal}.  
Therefore $H_r=G^\star$.

The final statement follows immediately: if a parametric family $\{\pi_\phi\}$ contains $G^\star$ and the RL algorithm converges to a global maximizer of $J(\pi_\phi)$, then the limit point must be $G^\star$.

\noindent\textbf{Proof of Theorem~\ref{thm:sampling-optimal}.} For any measurable set $A\subseteq\mathcal{X}$, consider a single proposal step of
Algorithm~\ref{alg:two_rate_sampling_pvalue}.  
A proposal $x$ is drawn from $F$, and then accepted with probability
\[
a(x)
=
\begin{cases}
1, & x\in S,\\[3pt]
w_0/w_1, & x\in S^c.
\end{cases}
\]
Thus the joint probability that the algorithm proposes a point in $A$ and accepts it in this step is
\[
\Pr(x\in A,\mathrm{acc})
=
F(A\cap S) + \frac{w_0}{w_1}F(A\cap S^c).
\]

The overall acceptance probability in one proposal step is
\[
p_{\mathrm{acc}}
=
\mathbb{E}_{X\sim F}[a(X)]
=
F(S)\cdot 1 + F(S^c)\cdot \frac{w_0}{w_1}
=
\alpha + (1-\alpha)\frac{w_0}{w_1}.
\]
Multiplying numerator and denominator by $w_1$ gives
\[
p_{\mathrm{acc}}
=
\frac{w_1\alpha + w_0(1-\alpha)}{w_1}.
\]
By the definitions $w_1=(1-\beta)/\alpha$ and $w_0=\beta/(1-\alpha)$, we have
\[
w_1\alpha = 1-\beta,
\qquad
w_0(1-\alpha) = \beta,
\]
so
\[
w_1\alpha + w_0(1-\alpha) = (1-\beta)+\beta = 1,
\]
and therefore
\[
p_{\mathrm{acc}} = \frac{1}{w_1}.
\]

The distribution of an accepted proposal in a single step is the proposal distribution
conditioned on acceptance:
\[
\Pr(x\in A\mid\mathrm{acc})
=
\frac{\Pr(x\in A,\mathrm{acc})}{p_{\mathrm{acc}}}
=
\frac{F(A\cap S) + (w_0/w_1)F(A\cap S^c)}{1/w_1}
=
w_1 F(A\cap S) + w_0 F(A\cap S^c).
\]
Since successive proposal steps are independent and each accepted sample is obtained
by repeating this mechanism until an acceptance occurs, the law of the finally accepted
sample is exactly $\Pr(x\in A\mid\mathrm{acc})$.
Thus the accepted sample has distribution
\[
G^\star(A)
=
w_1 F(A\cap S) + w_0 F(A\cap S^c)
=
\frac{1-\beta}{\alpha}F(A\cap S)
+
\frac{\beta}{1-\alpha}F(A\cap S^c),
\]
which matches the optimal watermarked distribution characterized in Theorem~\ref{thm:f-optimal}.

\section{Additional Discussions on Robustness}
Our analysis begins with a detector that satisfies the Type~I and
Type~II constraints~\eqref{eq:FPR} and~\eqref{eq:FNR}, and focuses on the
subsequent question of how to construct the optimal watermarked
distribution once such a detector is fixed.  Under this assumption, our
results characterize both the minimal fidelity loss required for
detectability and the optimal distribution that attains this optimum.
The design of the detector itself is not the subject of this work, and
none of our results depend on any particular choice of $D$.  

In practical watermarking systems, robustness depends heavily on the
detector itself. Different choices of $D$ lead to different robustness behaviours.
For instance, when watermarking large language models, one may adopt a token-level detector that flags a sample whenever the proportion of tokens falling into a predefined subset exceeds a threshold (\cite{kirchenbauer2023watermark}) or a
more semantic detector based on higher-level features of the generated
text (\cite{mitchell2023detectgpt}).  Such design choices influence the system's resistance to attacks,
but they are orthogonal to the questions addressed here.

The focus of this work is complementary: given {any} detector that
meets the desired false-positive constraint, we characterize the optimal
watermarked distribution associated with it and the inherent fidelity
cost required for detectability. In this sense, detector design and
watermark construction play distinct roles. Our results apply to the
latter, and remain valid regardless of how the former is implemented.









\end{document}